\documentclass[fleqn, 10pt]{wlscirep}

\usepackage{multicol, multirow}
\usepackage{graphicx}
\usepackage{float}

\title{Tailoring the Spectral Absorption Coefficient of a Blended Plasmonic Nanofluid Using a Customized Genetic Algorithm}

\author[1,2]{Junyong Seo}
\author[1,2]{Caiyan Qin}
\author[1,2]{Jungchul Lee}
\author[1,2,*]{Bong Jae Lee}

\affil[1]{Department of Mechanical Engineering, Korea Advanced Institute of Science and Technology, Daejeon 34141, South Korea}
\affil[2]{Center for Extreme Thermal Physics and Manufacturing, Korea Advanced Institute of Science and Technology, Daejeon 34141, South Korea}
\affil[*]{Corresponding author: bongjae.lee@kaist.ac.kr (B.J. Lee)}

\begin{abstract}
Recently, plasmonic nanofluids (i.e., a suspension of plasmonic nanoparticles in a base fluid) have been widely employed in direct-absorption solar collectors because the localized surface plasmon supported by plasmonic nanoparticles can greatly improve the direct solar thermal conversion performance. Considering that the surface plasmon resonance frequency of metallic nanoparticles, such as gold, silver, and aluminum, is usually located in the ultraviolet to visible range, the absorption coefficient of a plasmonic nanofluid must be spectrally tuned for full utilization of the solar radiation in a broad spectrum. In the present study, a modern design process in the form of a genetic algorithm (GA) is applied to the tailoring of the spectral absorption coefficient of a plasmonic nanofluid. To do this, the major components of a conventional GA, such as the gene description, fitness function for the evaluation, crossover, and mutation function, are modified to be suitable for the inverse problem of tailoring the spectral absorption coefficient of a plasmonic nanofluid. By applying the customized GA, we obtained an optimal combination for a blended nanofluid with the desired spectral distribution of the absorption coefficient, specifically a uniform distribution, solar-spectrum-like distribution, and a step-function-like distribution. The resulting absorption coefficient of the designed plasmonic nanofluid is in good agreement with the prescribed spectral distribution within about 10\% to 20\% of error when six types of nanoparticles are blended. Finally, we also investigate how the inhomogeneous broadening effect caused by the fabrication uncertainty of the nanoparticles changes their optimal combination. 
\end{abstract}

\begin{document}

\maketitle

\section{Introduction}
Plasmonic nanofluids, which contain a suspension of plasmonic nanoparticles in a base fluid, have been proposed as effective working fluids to directly convert solar radiation to thermal energy \cite{lee2012radiative}. Owing to the resonance characteristics of the localized surface plasmon (LSP), the absorption efficiency of the nanoparticles can be greatly enhanced with the excitation of the LSP, offering great potential in solar thermal applications. For instance, a direct-absorption solar collector (DASC) combined with a plasmonic nanofluid has drawn much attention for solar thermal energy harvesting in recent decades \cite{lee2012radiative, jeon2014optical, duan2018photothermal, mallah2018blended, mehrali2018full}. Recently, Qin \textit{et al.} \cite{qin2018optimization} showed how the spectral absorption coefficient of a plasmonic nanoparticle should be tuned (i.e., either uniformly or following the solar spectrum) by engineering nanoparticle suspensions to exploit the solar radiation maximally with the given constraint of the total particle concentration. Therefore, the effective tuning of the spectral absorption coefficients of plasmonic nanofluids is crucial for improving the thermal performance capabilities of DASCs. 

As suggested by Lee \textit{et al.} \cite{lee2012radiative}, broadband absorption spectra can be designed by blending multiple types of nanoparticles given that the resonance wavelength of the LSP depends on the material, size and shape of the nanoparticles. The simplest and the most common structure is a spherical nanoparticle \cite{encina2010plasmon, schaeublin2012does}. Nano-spheres made with noble metals are widely utilized in various disciplines, such as in medical \cite{huang2010gold} and biological \cite{murphy2008gold, schaeublin2012does} applications. However, because the resonance peak of the LSP associated with nano-spheres mainly depends on the material properties \cite{maier2007plasmonics}, nano-spheres themselves may not be suitable for thermal applications. Thus, additional types of nanoparticles, such as silica core-metallic shell \cite{khanadeev2017optical, ma2017controllable} and nano-rod \cite{yu1997gold, schaeublin2012does} nanoparticles, have also been considered given their potential for better controllability. For core-shell nanoparticles, the ratio between the core radius and the shell thickness serves as a factor when tuning the absorption response \cite{oldenburg1998nanoengineering}, while the aspect ratio of the nano-rod performs this function \cite{link1999spectral}. Thus, the critical question is \textit{``What would the optimal combination of various types of nanoparticles be for the effective tuning of the spectral absorption coefficient?''} Note that finding the optimal combination of plasmonic nanoparticles is not a straightforward task due to the diversity and complexity of nanoparticles with regards to their materials and shapes. For instance, Taylor \textit{et al.} \cite{taylor2012nanofluid} just employed the Monte-Carlo approach (i.e., random generation and selection) to find the optimal blending combination of core-shell particles for a nanofluid-based optical filter.

In general, an inverse problem is a problem that requires the determination of the design of a system from its output response. It is known that finding a proper solution to an inverse problem is often challenging because most inverse problems are ill-posed and nonlinear \cite{modest2013radiative}. Nevertheless, if a particular solution-finding technique of an inverse problem is available, it can be readily applied to diverse engineering fields, such as magnetic resonance imaging \cite{sumpf2011model}, combustion \cite{liu2019inverse}, and radiative heat transfer \cite{yadav2019inverse, radfar2019application}. Tailoring the absorption spectrum of a plasmonic nanofluid can also be treated as an inverse problem when designing a system (i.e., combination of plasmonic nanoparticles) at a given response (i.e., the desired spectral absorption coefficient). Because a blended combination of nanoparticles is represented with a broad range of variables, it is difficult to determine the optimal composition of a nanofluid to have the desired absorption spectrum. Therefore, a modern solution technique, such as a genetic algorithm \cite{yadav2019inverse, radfar2019application}, can be employed to solve our blending problem.

Genetic algorithms (GAs) have been widely employed to solve many design problems by customizing a description of an individual chromosome (i.e., member of a population) and a fitness function (i.e., score of the chromosome) properly \cite{goldberg1988genetic, harman2001search}. For example, the dimensions of a tandem-grating nanostructure for a solar thermal absorber \cite{choi2018robust}, the spectral distribution of absorption coefficients for DASC \cite{qin2018optimization}, and the dimensions of a multi-layer microcylinder for a plasmonic nanojet \cite{huang2019optimization} have been optimized based on carefully defined chromosomes and fitness functions. Here, we also apply a GA to find the optimal combination of plasmonic nanoparticles to achieve the desired spectral absorption coefficient of a nanofluid. To maximize the diversity of the plasmonic response of the nanoparticles, we consider two materials (gold and silver) and three types of nanoparticle shapes (nano-sphere, core-shell, and nano-rod). The target spectral absorption coefficient of the plasmonic nanofluid is first set to be either uniform or to follow the solar spectrum \cite{qin2018optimization}. In addition, a step-function-like absorption coefficient will be designed for a hybrid solar PV/T application \cite{taylor2012nanofluid, jia2019development}. Finally, how the inhomogeneous broadening effect caused by the fabrication uncertainty of the nanoparticles \cite{khanadeev2017optical, jeon2014optical} changes their optimal combination is also investigated.

\section{Modelling}
\subsection{Absorption coefficient of a blended plasmonic nanofluid}

\begin{figure}[!t]
    \centering
    \includegraphics[width=0.75\textwidth]{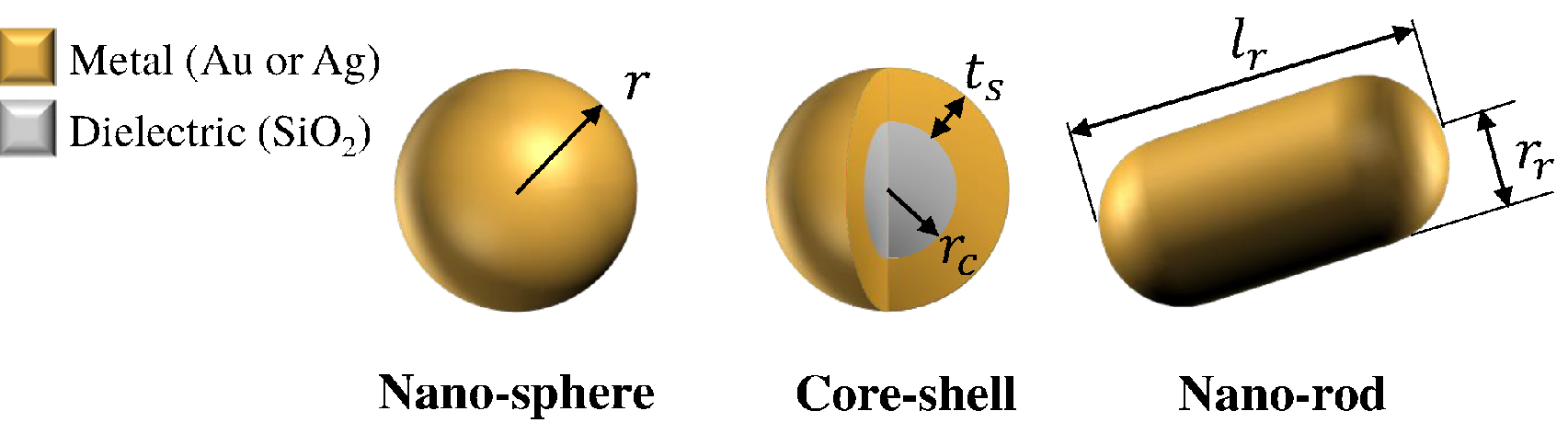}
    \caption{Schematic of the nano-sphere, core-shell, and nano-rod shapes. The design variables are the radius ($r$) for the nano-sphere, the core radius ($r_c$) and the shell thickness ($t_s$) for the core-shell, and the radius ($r_r$) and the length ($l_r$) for the nano-rod.}
    \label{fig:particleScheme}
\end{figure}

It is well known that subwavelength-size metallic nanoparticles can support a localized surface plasmon (LSP), whose resonance condition depends strongly on the materials, sizes, and shapes of the nanoparticles \cite{bohren1983absorption, hutter2004exploitation, jain2006calculated}.
In the present study, we consider two materials and three shapes of nanoparticles (see Fig.\ \ref{fig:particleScheme}) to diversify a number of possible blending combinations. The ranges of each design variable are carefully constrained according to the literatures \ \cite{khanadeev2017optical, koktan2017magnetic, jeon2014optical}. These are listed in Table\ \ref{tbl:variableRange}.
\begin{table}[!b]
    \centering
    \caption{Ranges of each design variable when training the surrogate model}
    \begin{tabular}{c|c|c}
    \hline
    Type & Design variable & Range (nm) \\ \hline
    Nano-sphere & Radius ($r$) & 10 $\sim$ 100 \\ \hline
    \multirow{2}{*}{Core-shell} & Core radius ($r_c$) & 5 $\sim$ 90\\
    & Shell thickness ($t_s$) & 5 $\sim$ ($100 - r_c$)\\ \hline
    \multirow{2}{*}{Nano-rod} & Radius ($r_r$) & 6 $\sim$ 30\\
    & Length ($l_r$) & $\text{max}(16, 2r_r)$ $\sim$ 200\\ \hline
    \end{tabular}
    \label{tbl:variableRange}
\end{table}

For a given nanoparticle, its spectral absorption efficiency, $Q_{a}(\lambda)$, can be calculated by solving Maxwell's equations. For nano-sphere and core-shell particles, Mie-scattering theory \cite{bohren1983absorption} and a modified version of it \cite{lee2012radiative} were used to determine $Q_{a}(\lambda)$. For the nano-rod, a boundary element method (BEM) was applied to obtain the polarization- and direction-averaged absorption efficiency. In this study, the open-source BEM software MNPBEM\cite{waxenegger2015plasmonics} was used. In the calculations, the permittivities of gold, silver, silicon dioxide, and water (i.e., the base fluid) were used from tabulated data \cite{palik1998handbook}. As discussed by Lee \textit{et al.}\cite{lee2012radiative}, if the radius (or thickness) of a metal is smaller than its mean-free-path of conduction electrons, we also must consider that the size-dependent permittivity of the metal should differ from that of the bulk metal due to electron-boundary scattering. Furthermore, a broadening effect will arise due to the modification of the permittivity. In this work, the effect of electron-boundary scattering is neglected for simplicity, though this effect will be discussed later.

With the calculated absorption efficiency of the $i$-th particle in water, $Q_{a, i}(\lambda)$, the corresponding absorption coefficient of the nanofluid, $\alpha_{a, \lambda}$, can be expressed as\cite{bohren1983absorption}:
\begin{equation}
    \alpha_{i, \lambda} = \frac{3 f_i}{2 D_i}Q_{a, i}(\lambda)\label{eq:eachAbsorptionCoefficient}
\end{equation}
where, $f_i$ is the volume concentration of the nanoparticle and $D_i = \sqrt[3]{6\times\text{(volume of particle)}/\pi}$ is the effective diameter of each particle \cite{jeon2014optical}. If $N$ types of nanoparticles are dispersed together, the resulting absorption coefficient of blended plasmonic nanofluid ($\alpha_\lambda$) is then given by: 
\begin{equation}
    \alpha_\lambda = \left ( 1 - \sum^N_{i=1}{f_i} \right ) \alpha_{w, \lambda} + \sum^N_{i=1}{\alpha_{i, \lambda}}\label{eq:absorptionCoeffcient}
\end{equation}
where, $\alpha_{w, \lambda}$ is the absorption coefficient of the water itself. Because solar irradiance begins at approximately $\lambda= 300$ nm and the absorption coefficient of water becomes dominant after $\lambda= 1,100$ nm, we calculate the absorption efficiency spectrum of each particle from 300 nm to 1,100 nm in 10 nm intervals (i.e., 81 spectral data points).

In principle, $Q_{a, i}(\lambda)$ must be known \textit{a priori} in each computation of the fitness function of a GA. Because the calculation of $Q_{a, i}(\lambda)$ takes about 3 min and the average number of fitness calculations in our GA numbers into the thousands, it is not feasible to compute it every time. To reduce the computational cost, we decided to build a surrogate model to estimate the absorption efficiency of each nanoparticle, i.e., [Input: geometry of $i$-th particle and $\lambda$ $\rightarrow$ Output: $Q_{a, i}(\lambda)$]. To ensure the accuracy of the surrogate model, an artificial neural network model\cite{seo2019design} was employed as part of a modelling technique. To train the neural network models, samples were composed with 2 nm intervals of the design variables and a 10 nm interval of the wavelength. Consequently, we constructed and applied accurate surrogate models with correlation values exceeding $R=0.999$ with a regulated amount of sample data. Note that neural networks were modelled with three fully connected hidden layers and 10 nodes in each layer, which was enough to construct a surrogate model for estimating the $Q_{a}$ spectrum. The accuracy of the model was estimated with the difference between the predicted and actual $Q_{a, i}(\lambda)$ values at the peak location, where the maximum $Q_{a, i}$ value was achieved. As a result, the accuracy of model used in this work was found to be between 0.3\% (for the nano-spheres) and 1.6\% (for the core-shells and the nano-rods) on average.

\subsection{Customized genetic algorithm}
The genetic algorithm (GA) is a powerful method for solution processes owing to its ranges for diverse applicability for a variety of problems \cite{goldberg1988genetic, harman2001search}. In the world of a GA, the population consists of individuals. Each individual has its own chromosome (i.e., set of genes) and evolves along descent generations. Based on simple and bio-mimicking procedures, the population of the GA will evolve to obtain the best individual, which has the best chromosome, through a process of selection, crossover and mutation. The GA can be utilized with proper modification of its gene description, fitness function, and any embedded algorithms or hyper-parameters. In this study, (1) descriptions of the chromosomes (or genes), (2) fitness function, (3) crossover, and (4) mutation algorithms are customized especially for solving our inverse problem, designing of the system (a combination of plasmonic nanoparticles) at the given response (the desired spectral absorption coefficient).

The most significant aspect of customization is to define the chromosomes of the GA. Because the chromosomes of individuals must be related to a combination of plasmonic nanoparticles, we defined the chromosome to possess a set of nanoparticles as a genes. Each gene on the chromosome has particle properties, such as the material (gold or silver), shape (nano-sphere, core-shell, nano-rod), design variables (geometric parameters), and volume concentration. A chromosome is implemented as a list of genes. When a chromosome is created, the properties of each gene are determined randomly within their ranges. Note that the volume concentration is intentionally set to be less than 0.005\% divided by the number of genes (i.e., the number of nanoparticle types in Eq.\ \eqref{eq:absorptionCoeffcient}) to match the scale of each particle's volume fraction to 0.0001\% \cite{jeon2014optical}.

A fitness function is usually set to be a distance or a loss function, as the optimization process evolves to minimize a score. In this work, the fitness function is defined as the sum of square error (SSE):
\begin{equation}
    \text{SSE} = \sum_{j=0}^{80}{(\alpha_{\lambda_j} - \alpha_{\text{target}, \lambda_j})^2}
\end{equation}
where, $\lambda_i$ is the wavelength in interest with a 10 nm interval (i.e., $300 + 10j$ nm) and $\alpha_{\text{target}, \lambda}$ is the target absorption coefficient spectrum defined in Section\ \ref{sec:targetAbs}. The absorption coefficient of the blended nanofluid was calculated from Eqs.\ \eqref{eq:eachAbsorptionCoefficient} and \eqref{eq:absorptionCoeffcient}. To obtain $Q_{a, i}(\lambda)$, the type and design variables of the nanoparticles described in the chromosome are required. The GA scored each individual with this SSE value and caused the population to evolve to minimize the score.

After the scores of individuals are evaluated by the fitness function, the GA will prepare the population of the next generation. Initially, the highest scoring individual will remain based on elitism. By default, the top 5\% individuals in terms of their scores will move to the next generation. Next, the GA selects individuals as the parents of the rest (i.e., 95\% of the next generation) according to rule of natural selection. That is, individuals with better fitness values are more probable to be a parent, and a stochastic uniform selection rule \cite{baker1987reducing} is applied. For the crossover process, an offspring individual will have a gene list, which basically consists of the first parent's genes. In addition, an arbitrary gene fragment cut from the second parent is inserted into a randomly chosen location. Finally, for the mutation process, simple one-point mutation is used. A mutated child will have a gene list from a parent with one point of a gene replaced by newly created nanoparticle. In this work, a 20\% mutation rate is used by default. In other words, 80\% of the remaining children are generated by crossover while 20\% are generated by mutation. Hence, 162\% ($95\% \times [80\times2 + 20]\%$) of the total population is selected by the selection rule, and the offspring for the next generation is born from them by following the crossover and mutation process.

\begin{figure}[!t]
    \centering
    \includegraphics[width=.99\textwidth]{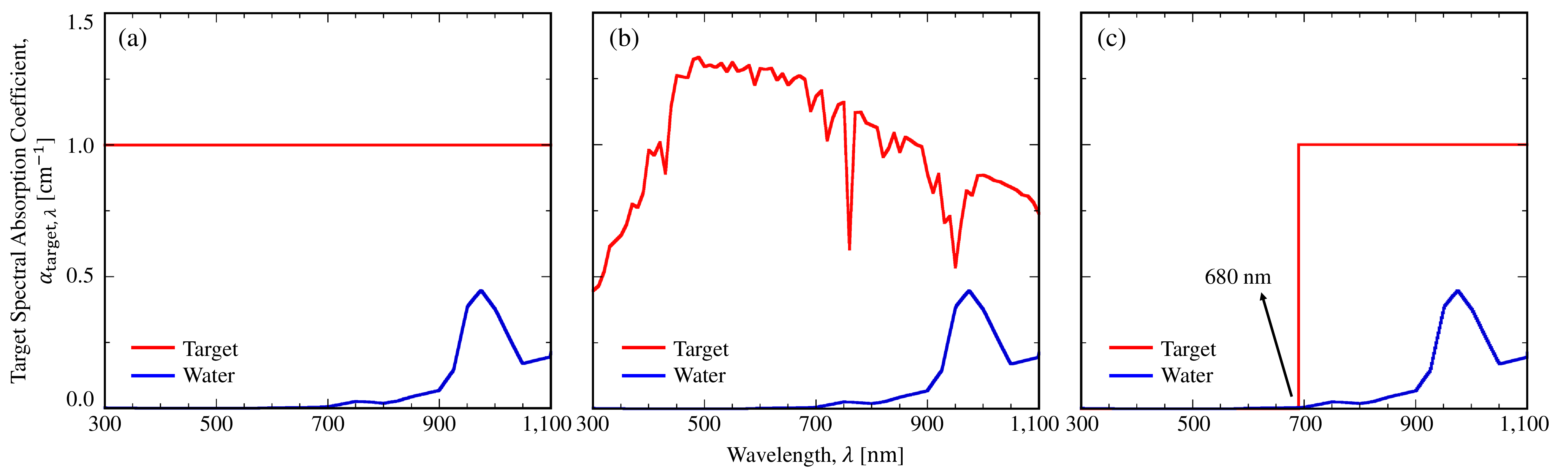}
    \caption{Target absorption coefficient of the blended plasmonic nanofluid (red line): (a) uniform distribution, (b) solar-spectrum-like distribution, and (c) step-function-like distribution. The spectral absorption coefficient of water is also illustrated by the blue line.}
    \label{fig:targetAbs}
\end{figure}
%
\subsection{Target spectrum of the absorption coefficient} \label{sec:targetAbs}
To demonstrate how the customized GA can effectively tune the absorption coefficient of a blended plasmonic nanofluid, we consider three target absorption coefficient spectra, as illustrated in Fig.\ \ref{fig:targetAbs}. The first two target spectra are for solar thermal applications, especially for a direct-absorption solar collector, i.e., a uniform distribution (Fig.\ \ref{fig:targetAbs}a) and a solar-spectrum-like distribution (Fig.\ \ref{fig:targetAbs}b). As reported by Qin \textit{et al.} \cite{qin2018optimization}, a uniform absorption coefficient is more efficient for a highly concentrated nanofluid because the heat loss can be minimized. On the other hand, a solar-spectrum-like absorption coefficient is more suitable when the system only requires an insufficient particle concentration. For simplicity, the average of the absorption coefficient was set to 1 cm$^{-1}$ considering that the magnitude of the absorption coefficient is scalable according to the particle concentration [see Eq.\ \eqref{eq:absorptionCoeffcient}]. It should be noted from Fig.\ \ref{fig:targetAbs} that the absorption coefficient of water does not play much of a role in the absorption process.
Thus, the nanoparticles should be carefully designed to achieve broadband absorption, associated with their LSP resonances.
In addition, a step-function-like absorption coefficient (Fig.\ \ref{fig:targetAbs}c) is designed for hybrid solar photovoltaic/thermal (PV/T) applications \cite{jia2019development, taylor2012nanofluid}. At the zero-absorption regime of the nanofluid, incident solar irradiance will directly reach the PV cell and be converted into electricity. In the opaque regime of the nanofluid, the solar irradiance will be converted to heat by the nanofluid. Here, we select the step point of the spectrum to be the bandgap of the PV cell used in a hybrid PV/T system with a high bandgap with 1.84 eV (approximately 680 nm) \cite{abdulraheem2014optical}. Note also that a high-bandgap PV cell is widely applied for common solar cell systems \cite{todorov2017ultrathin} or for special purposes \cite{jenkins2013high}.

\section{Results and Discussion}
The main idea when tailoring the absorption coefficient of a blended plasmonic nanofluid is to distribute the absorption peaks associated with each type of nanoparticle along the target spectrum. It is thus expected that more types of nanoparticles makes the corresponding absorption coefficient a better fit to the target spectrum. Considering the productivity of a plasmonic nanofluid, the number of types of nanoparticles ($N$) should not be excessive. Given that plasmonic nanofluids with 3 to 5 types of nanoparticles have been experimentally demonstrated \cite{jeon2014optical, mallah2018blended}, $N$ is limited to 6 or less. 

Although not shown here, a higher value of $N$ can achieve a smaller root-mean-square error (RMSE), defined as $\text{RMSE}=\sqrt{\text{SSE}/81}$. Henceforth, we discuss the optimal blending combination case of $N=6$, which can retain the smallest RMSE value (i.e., the closest absorption coefficient spectrum to the target). The detailed dimensions as well as the locations of the major absorption peaks of each type of nanoparticles are listed in Table\ \ref{tbl:blendingResult}.

\begin{table}[!t]
    \centering
    \caption{Optimal combination of plasmonic nanoparticles for the desired spectral absorption coefficient of the nanofluid. The design variables are $r$ for the nano-sphere, ($r_c$, $t_s$) for the core-shell, and ($r_r$, $l_r$) for the nano-rod.}
    \begin{tabular}{c | c | c | c | c}
    \hline
    Particle type & \multirow{2}{*}{Property} & Uniform & Solar-spectrum-like & Step-function-like\\ 
    Index & & distribution & distribution & distribution \\ \hline
    \multirow{5}{*}{\#1} & Material & Silver & Gold & Silver \\
    & Type & Core-shell & Core-shell & Core-shell \\
    & Design & (73, 14) & (15, 11) & (86, 8) \\
    & $f_i\times 10^{6}$ & 1.053 & 0.544 & 3.870 \\
    & Peak [nm] & 580 & 550 & 720 \\ \hline
    \multirow{5}{*}{\#2} & Material & Silver & Silver & Silver \\
    & Type & Core-shell & Nano-rod & Core-shell \\
    & Design & (80, 9) & (16, 120) & (87, 5) \\
    & $f_i\times 10^{6}$ & 3.602 & 0.187 & 1.547 \\
    & Peak [nm] & 670 & 380, 810 & 850 \\ \hline
    \multirow{5}{*}{\#3} & Material & Gold & Silver & Gold \\
    & Type & Nano-rod & Core-shell & Core-shell \\
    & Design & (28, 181) & (49, 13) & (86, 6) \\
    & $f_i\times 10^{6}$ & 0.605 & 4.224 & 0.036 \\
    & Peak [nm] & 510, 890 & 500, 690 & 810 \\ \hline
    \multirow{5}{*}{\#4} & Material & Gold & Silver & Silver \\
    & Type & Core-shell & Core-shell & Core-shell \\
    & Design & (52, 12) & (88, 10) & (89, 7) \\
    & $f_i\times 10^{6}$ & 3.116 & 2.642 & 1.933 \\
    & Peak [nm] & 610, 740 & 680 & 760 \\ \hline
    \multirow{5}{*}{\#5} & Material & Gold & Silver & Gold \\
    & Type & Core-shell & Core-shell & Core-shell \\
    & Design & (83, 6) & (53, 7) & (46, 5) \\
    & $f_i\times 10^{6}$ & 1.298 & 0.970 & 0.599 \\
    & Peak [nm] & 800, 1080 & 610, 820 & 860 \\ \hline
    \multirow{5}{*}{\#6} & Material & Gold & Gold & Silver \\
    & Type & Core-shell & Core-shell & Core-shell \\
    & Design & (88, 5) & (87, 5) & (15, 70) \\
    & $f_i\times 10^{6}$ & 1.434 & 1.909 & 0.003 \\
    & Peak [nm] & 860 & 860 & 400 \\ \hline
    \end{tabular}
    \label{tbl:blendingResult}
\end{table}
\begin{figure}[!t]
    \centering
    \includegraphics[width=.55\textwidth]{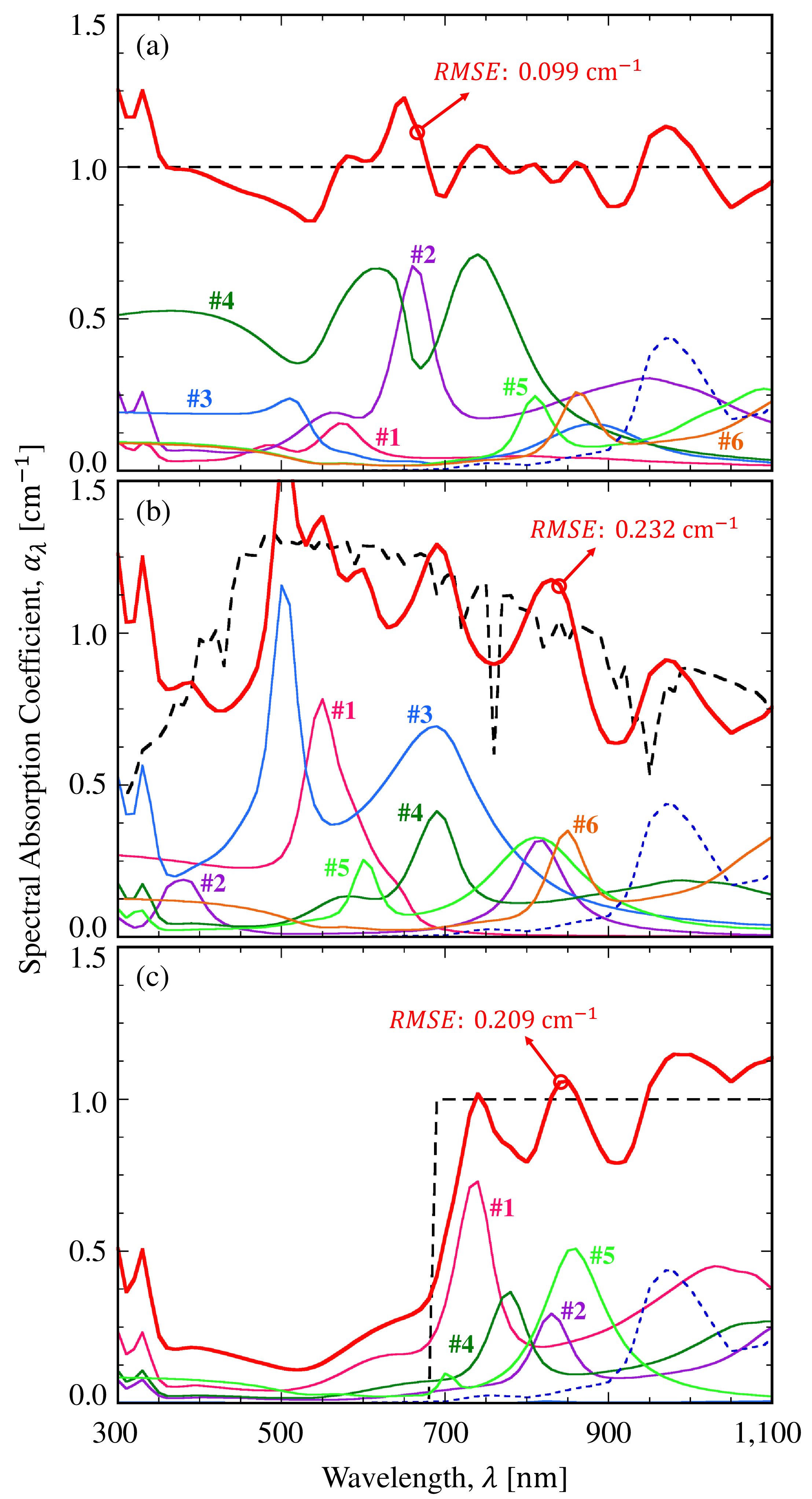}
    \caption{Absorption coefficient of a blended plasmonic nanofluid for the target spectrum: (a) uniform distribution, (b) solar-spectrum-like distribution, and (c) step-function-like distribution. The effect of water is illustrated by the blue dotted line.}
    \label{fig:blendingResult}
\end{figure}

Figure \ref{fig:blendingResult}a shows the absorption coefficient of a blended plasmonic nanofluid for the target spectrum with a uniform distribution. The designed plasmonic nanofluid results in a RMSE value of 0.099 cm$^{-1}$, which is less than 10\% of the target absorption coefficient (i.e., 1 cm$^{-1}$). The absorption peaks of each type of nanoparticle are well distributed along the visible and near-infrared spectral regions for broadband absorption. Although the \#4 particle (i.e., the Au core-shell) mainly contributes to the visible absorption, the \#2 and \#3 particles successfully compensate for the absorption dips of the \#4 particle. Moreover, all six types of nanoparticles shows minor absorption at the wavelengths greater than 900 nm, where water (i.e., the base fluid) starts to play a role. It is interesting to note that no nano-spheres are used in the six types of nanoparticles. This is mainly due to the tunability of the LSP resonance condition in the core-shell and nano-rod structures. In other words, the nano-sphere is less tunable as the polarizability of its Clausius-Mossotti relation is predetermined by the material properties \cite{bohren1983absorption} and the resulting absorption peak is often confined to a narrow spectral range (i.e., Au: $500 \sim 540$ nm and Ag: $380 \sim 450$ nm). 

The absorption coefficient of a blended plasmonic nanofluid for a solar-spectrum-like target spectrum is shown in Fig.\ \ref{fig:blendingResult}b. To follow the solar spectrum, the absorption peaks should be confined to the major spectral regions of solar radiation (i.e., from 400 to 700 nm and from 800 to 900 nm). It can be observed that the \#1, \#3, and \#4 particles mainly contribute to the absorption in the aforementioned spectral region. Although the RMSE value is 0.232 cm$^{-1}$ (about twice that the uniform case) for the solar-spectrum-like distribution, the designed absorption coefficient reasonably follows the solar spectrum except for wavelengths between 300 and 400 nm. 

Finally, we also demonstrate the blended plasmonic nanofluid for the step-function-like distribution in Fig.\ \ref{fig:blendingResult}c.
As in the case of the solar-spectrum-like distribution, the designed absorption coefficient captures the features of the target spectrum. However, there exists non-negligible absorption of approximately 0.2 cm$^{-1}$ in the wavelengths between 300 and 680 nm, mainly due to intrinsic absorption by silver and gold. The resulting RMSE value is 0.209 cm$^{-1}$, which is slightly less than that in Fig.\ \ref{fig:blendingResult}b. Interestingly, Table \ref{tbl:blendingResult} reveals that the volume fractions of the \#4 and \#6 particles were greatly suppressed by the GA, meaning that the \#4 and \#6 particles were considered to be "useless" by the GA. Therefore, we can simply use four types of particles for the step-function-like distribution without seriously compromising the fitness. In Fig.\ \ref{fig:blendingResult}, the customized GA clearly demonstrates its excellent design capability for a blended plasmonic nanofluid with the desired spectral absorption coefficient. 

Table\ \ref{tbl:blendingResult} also indicates that only core-shell nanoparticles are used for the step-function-like distribution. Although the nano-rod can induce the multiple LSP peaks \cite{yu1997gold}, its resonance condition is polarization-dependent due to its geometrical anisotropy. Hence, the polarization-averaged absorption efficiency becomes less significant as compared to the geometrically isotropic core-shell structure. The present optimization results clearly indicate that the core-shell nanoparticle is superior to the nano-sphere and the nano-rod structures in terms of the tunability of the LSP resonance condition as well as the enhanced absorption efficiency associated with the LSP. 

\begin{figure}[!b]
    \centering
    \includegraphics[width=0.55\textwidth]{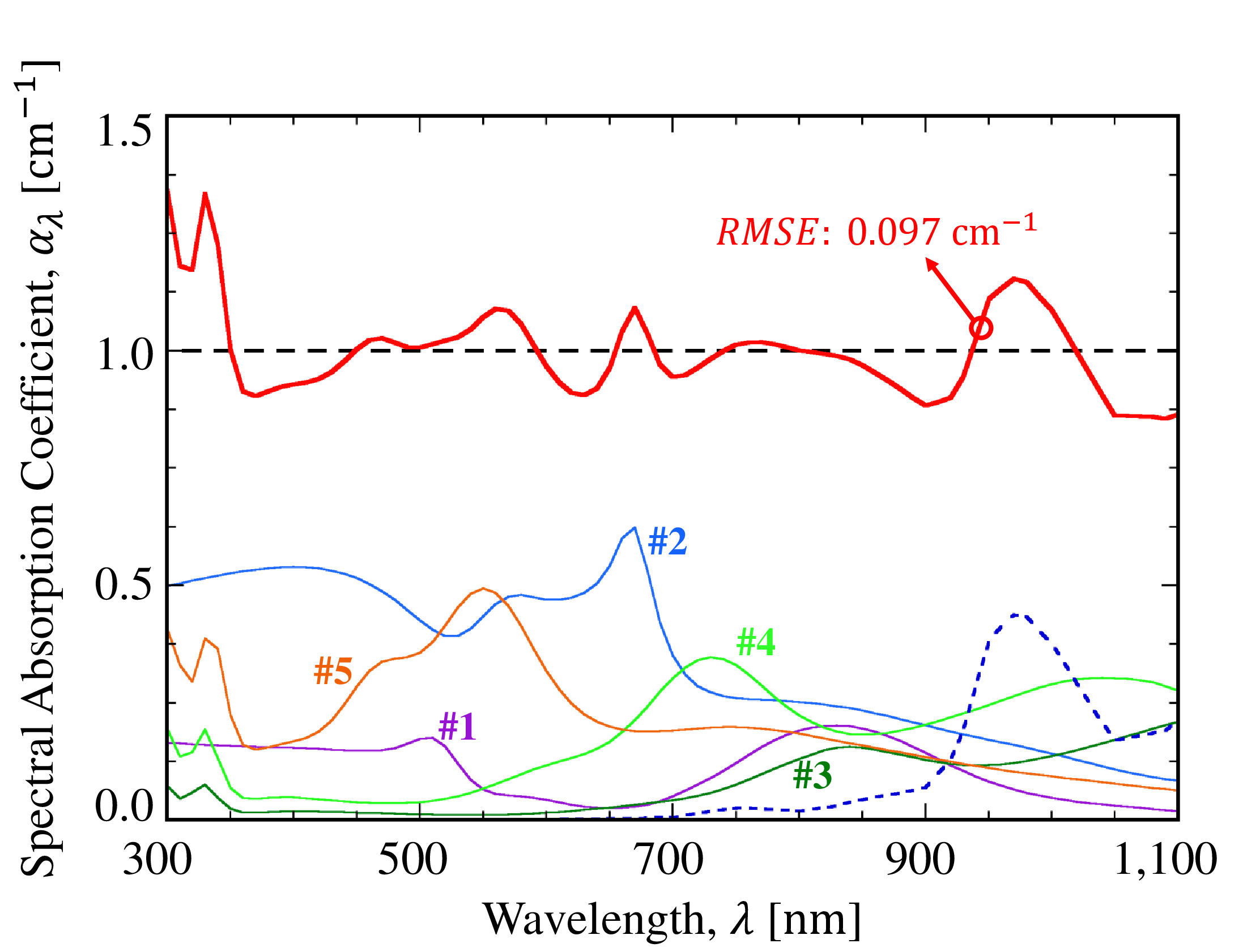}
    \caption{Absorption coefficient of a blended plasmonic nanofluid considering inhomogeneous broadening due to polydispersed nanoparticles in reality. The effect of water is also illustrated by the blue dotted line.}
    \label{fig:broadeningResult}
\end{figure}
\begin{table}[!t]
    \centering
    \caption{Optimal combination of plasmonic nanoparticles considering inhomogeneous broadening due to polydispersed nanoparticles in reality. The notation of the design variables follows that in Table\ \ref{tbl:blendingResult}.}
    \begin{tabular}{c | c c c c c}
    \hline
    Property & \#1 &  \#2 &  \#3 &  \#4 &  \#5 \\ \hline
    Material & Gold & Gold & Silver & Silver & Silver \\
    Type & Nano-rod & Core-shell & Core-shell & Core-shell & Core-shell\\
    Design & (21, 139) & (76, 15) & (88, 5) & (89, 8) & (71, 15) \\
    $f_i\times 10^{6}$ & 0.41 & 4.84 & 1.50 & 3.23 & 4.28 \\ 
    Peak [nm] & 510, 830 & 670 & 840 & 740, 1050 & 560 \\ \hline
    \end{tabular}
    \label{tbl:broadeningResult}
\end{table}

Thus far, we have demonstrated how to achieve broadband absorption by blending nanoparticles made of noble metals (such as Au and Ag), which usually exhibits sharp resonance peaks. In reality, however, there could be many factors that give rise to a broadening effect, such as the electron-boundary scattering effect when the characteristic size of the metal is smaller than the mean-free-path of electrons or inhomogeneous broadening due to a non-uniform size distribution of the nanoparticles. Because the electron-boundary scattering effect occurs only for the core-shell structure with an extremely thin metallic shell \cite{lee2012radiative}, for instance, its effect may not be prominent as compared to the inhomogeneous broadening that occurs inevitably due to polydispersed nanoparticles \cite{khanadeev2017optical, jeon2014optical}. Here, we examine how inhomogeneous broadening occurring in reality can affect the optimal combination of plasmonic nanoparticles by applying a randomized distribution of design variables. To do this, a Gaussian distribution with the mean value of the design variable and a standard deviation of 10\% of the mean is assumed. For instance, the core radius of core-shell particle ($r_c$) is treated as a random variable following a normal distribution, $N(r_c, (0.1r_c)^2)$. In the calculation, 100 random particles following a Gaussian distribution were calculated using the surrogate neural network model and their absorption spectra were averaged to determine the broadened absorption spectrum of randomized nanoparticles.

Figure \ref{fig:broadeningResult} shows the absorption coefficient of a blended plasmonic nanofluid considering inhomogeneous broadening due to polydispersed nanoparticles in reality. It is remarkable that we can achieve an even lower RMSE value (i.e., 0.097 cm$^{-1}$) than the previous blending result ($\text{RMSE}=0.099$ cm$^{-1}$) with only five types of nanoparticles if inhomogeneous broadening is taken into account. The optimal combination of plasmonic nanoparticles considering inhomogeneous broadening is listed in Table\ \ref{tbl:broadeningResult}. As noted in Table\ \ref{tbl:broadeningResult}, the absorption peaks of each type of nanoparticle are well distributed, spanning the entire spectral region of interest, and the inhomogeneous broadening causes the absorption coefficient of the blended plasmonic nanofluid to be more uniform. Similarly, it is also expected that the electron-boundary scattering effect eventually makes the designed spectrum more uniform, possibly leading to a reduction in the required number of nanoparticle types. It should be noted that the optimal combination in Table \ref{tbl:broadeningResult} is wholly different from that in Table \ref{tbl:blendingResult}, suggesting that the customized GA is very effective at finding the solution under any given constraint.

\section{Summary}
We have employed a customized genetic algorithm to tailor the spectral absorption coefficient of a blended plasmonic nanofluid made of nano-sphere, core-shell, and/or nano-rod structures. The chromosome description, fitness function, crossover and mutation process in a conventional GA were customized to be suitable for the inverse problem of finding the optimal combination of plasmonic nanoparticles for the prescribed distribution of the absorption coefficient. In addition, neural network models estimating the absorption coefficient of a plasmonic nanoparticle were constructed and coupled with the customized GA to reduce the computational cost of the optimization process. In this work, three different target absorption coefficients, specifically a uniform distribution, solar-spectrum-like distribution and step-function-like distribution, were considered. The resulting absorption coefficient of a designed plasmonic nanofluid was in good agreement well with the prescribed spectral distribution within about 10\% to 20\% of error when six types of nanoparticles were used. Finally, we also considered inhomogeneous broadening mainly due to polydispersed nanoparticles during the optimization process. It was found that we can achieve an even lower RMSE value (i.e., 0.097 cm$^{-1}$) than in the previous blending result ($\text{RMSE}=0.099$ cm$^{-1}$) with fewer types of nanoparticles if inhomogeneous broadening is considered. The design methodology proposed here will facilitate the future development of a direct-absorption solar collector using a blended plasmonic nanofluid.

\section*{Data availability}

All data that support the findings of this study are available from the corresponding author upon request.

\bibliography{mainBib}

\section*{Acknowledgements}
This research was supported by the Basic Science Research Program (NRF-2019R1A2C2003605) and by the Creative Materials Discovery Program (NRF-2018M3D1A1058972) through the National Research Foundation of Korea (NRF) funded by the Ministry of Science and ICT. This research was also supported by the Korea Institute of Energy Technology Evaluation and Planning (KETEP) and the Ministry of Trade, Industry \& Energy (MOTIE) of the Republic of Korea (No.\ 20172010000850).

\section*{Author contributions}
Data preparation and customizing genetic algorithm were driven by J.S. under the supervision of B.J.L and J.L. The basic method for use of MNPBEM software was assisted by C.Q.

\section*{Competing interests}
The authors declare no competing financial and/or non-financial interests in relation to this work.

\end{document}